# Resolving the structure of the striped Ge layer on Ag(111): Ag$_2$Ge surface alloy with alternate fcc and hcp domains

K. Zhang,[1] D. Sciacca,[2] A. Coati,[3] R. Bernard,[1] Y. Borensztein,[1] P. Diener,[2] B. Grandidier,[2] I. Lefebvre,[2] M. Derivaz,[4] C. Pirri,[4] G. Prévot[1]

[1] Sorbonne Université, Centre National de la Recherche Scientifique, Institut des NanoSciences de Paris, INSP, F-75005 Paris, France

[2] Univ. Lille, CNRS, Centrale Lille, Univ. Polytechnique Hauts-de-France, Junia-ISEN, UMR 8520 - IEMN, F-59000 Lille, France

[3] Synchrotron SOLEIL, L'Orme des Merisiers Saint-Aubin, BP 48 91192 Gif-sur-Yvette Cedex, France

[4] Institut de Science des Matériaux de Mulhouse IS2M UMR 7361 CNRS-Université de Haute Alsace, 3 bis rue Alfred Werner, 68057, Mulhouse, France

**Abstract**

Two-dimensional (2D) honeycomb lattices beyond graphene, such as germanene, promise new physical properties such as quantum spin Hall effect. While there have been many claims of growth of germanene, the lack of precise structural characterization of the epitaxial layers synthesized hinders further research. The striped layer formed by Ge deposition on Ag(111) has been recently ascribed as a stretched germanene layer. Using surface X-ray diffraction and density functional theory calculations, we demonstrate that it corresponds in fact to a $Ag_2Ge$ surface alloy with an atomic density 6.45% higher than the Ag(111) atomic density. The overall structure is formed by stripes associated with a face-centered cubic top-layer alignment, alternating with stripes associated with an hexagonal-close-packed top-layer alignment, in great analogy with the $(22 \times \sqrt{3})$ Au(111) reconstruction.

**Introduction**

After the first experimental observation of the growth of Ge on Ag(111) in 1999 [1,2], the germanium/silver interface has been subject to a renewed interest since the discovery of germanene, the graphene-like allotrope for Ge atoms [3,4]. Ge structures formed upon deposition at room temperature on Ag(111) have been initially observed by low energy electron diffraction (LEED) and Auger electron spectroscopy (AES) [1]. Two ordered superstructures were found, corresponding to different Ge coverages. The first structure, corresponding to a $(\sqrt{3} \times \sqrt{3})R30°$ reconstruction related to the Ag(111) surface (hereafter named as $(\sqrt{3} \times \sqrt{3})$), was assigned to a Ge coverage $\theta_{Ge}$=1/3 ML, with respect to the atomic density of a Ag(111) plane. From core-level photoemission spectroscopy (XPS), a model of $Ag_2Ge$ surface alloy, where every three Ag atom is replaced with a Ge atom was proposed for the $(\sqrt{3} \times \sqrt{3})$ structure and confirmed from angle resolved photoemission spectroscopy (ARPES) [5]. The second structure, corresponding to a $(7 \times 7)$ reconstruction, was assumed to correspond to a complete Ge monolayer. This reconstruction was further recognized as a $c(\sqrt{3} \times 7)$ reconstruction and a model of Ge tetramers on top of a Ag(111) plane was proposed for its structure, based on scanning tunneling microscopy (STM) [6]. In contradiction with this model, from real-time STM observations of the growth, we have recently proposed that this latter structure is also a surface alloy with a coverage of $0.6 \pm 0.1$ ML [7].

The $(\sqrt{3} \times \sqrt{3})R30°$ structure obtained after deposition at room temperature was found to evolve upon annealing at 473 K [8]. A striped pattern, with ridges and valleys, appears in STM images, and LEED diffraction spots are split into satellite spots. Thus, this structure was shown to deviate from the simple $Ag_2Ge$ surface alloy and assigned to a $(6\sqrt{3} \times \sqrt{3})R30°$ reconstruction. This structure can also be obtained by evaporating Ge at 600K [9]. The striped pattern has been attributed to the relaxation of the compression force induced by the difference in the radii between Ge atoms and Ag atoms after Ge insertion [9].

The Ge/Ag(111) system has been recently revisited and contradictory observations have been reported. On the one hand, the striped phase has been interpreted as a highly stretched pure germanene layer, with 23% and 12% stretch in the direction parallel and perpendicular to the stripes [10]. On the other hand, in a recent study, the LEED diagram corresponding to satellite spots around diffraction conditions of a $(\sqrt{3} \times \sqrt{3})$ reconstruction, and previously attributed to the striped phase, has been interpreted as a $(19\sqrt{3} \times 19\sqrt{3})R30°$ reconstruction, corresponding to a $Ag_2Ge$ surface alloy contracted by 5% with respect to the Ag(111) surface [11].

In any event, in spite of several studies devoted to Ge/Ag(111), the precise atomic structure of the reconstruction formed at low coverage and for deposition above room temperature or after annealing is unknown. In particular, there is no consensus about the formation of a layer of germanene or of a Ag-Ge alloy. Surface X-ray diffraction (SXRD) is very well adapted for elucidating the structure of ordered surface reconstructions, and we have recently used it to successfully determine the precise atomic positions for the various silicene epitaxial layers on Ag(111) [12,13].

In this paper, we present a SXRD study of the $(\sqrt{3} \times \sqrt{3})$ Ge/Ag reconstruction. We show that the ordered structure associated with the satellite spots around $(\sqrt{3} \times \sqrt{3})$ diffraction conditions can be precisely indexed as a $c(31 \times \sqrt{3})$ reconstruction, and thus corresponds to the striped phase observed in STM images. Comparison with DFT calculations allows us to determine the precise atomic structure of the surface, which corresponds to a $Ag_2Ge$ surface alloy with an atomic density 6.45% higher than the Ag(111) atomic density. The overall structure is formed by stripes associated with a face-centered cubic (fcc) top-layer alignment, alternating with stripes associated with an hexagonal-close-packed (hcp) top-layer alignment, in great analogy with the $(22 \times \sqrt{3})$ Au(111) reconstruction.

**Methods**

SXRD experiments were performed at the SIXS beamline of SOLEIL synchrotron. The Ag(111) sample was prepared by repeated cycles of Ar+ sputtering and annealing at T=750 K. Ge was evaporated in the diffraction chamber from a crucible using a Knudsen cell with a sample kept at ∼ 420 K. The flux was estimated to 0.5 ML/h. The sample was analyzed with 18.46 keV X-rays at a grazing incidence angle of 0.2°. Scattered X-rays were detected with a X-ray Pixel chips with Adaptive Dynamics hybrid pixel detector [14]. Diffracted intensity was measured by performing a series of rocking scans around diffraction. We used the "BINoculars" software to produce three-dimensional (3D) intensity data in the reciprocal space from the raw data [15]. The intensity was further integrated along the direction parallel to the surface to obtain the structure factors. For this purpose, the data were fitted with the product of a lorentzian lineshape in one direction with a gaussian lineshape convolved with a door lineshape in the other direction, using a home-made software, and the fitted function has been analytically integrated. We finally obtained a set of 2493 structure factors along 62 inequivalent reconstruction rods. The $(h, k, l)$ indices used for indexing a reflection in reciprocal space refer to the Ag(111)) surface basis ($a_0 = b_0 = 2.889$ Å, $c_0 = 7.075$ Å, $\alpha_0 = \beta_0 = 90°$, $\gamma_0 = 120°$). The $(H, K, L)$ indices refer to the Ag(111) c(31 × √3) reconstruction basis ($a = 89.545$ Å, $b = 5.003$ Å, $c = 7.075$ Å, $\alpha = \beta = \gamma = 90°$).

Theoretical computations of the minimal energy configuration were done within the density functional theory (DFT). Calculations for the core electrons were performed using the projector augmented wave (PAW) method [16,17], as implemented in the VASP code [18,19]. The Perdew-Burke-Ernzerhof functional [20] was used to describe the exchange-correlation functional. The plane wave basis set was restricted to a cut-off of 300 eV. The Ag(111) substrate was modeled by a (31 × √3) nine-layer slab. Similar to the procedure adopted in ref. [5,13], the bottom two layers were kept fixed at the equilibrium theoretical positions. The 10[th] layer on top is modeled as a Ag$_2$Ge surface alloy following the SXRD results. The vacuum region is 9 Å thick. The entire system, Ag$_2$Ge alloy on silver, was fully relaxed by a conjugate gradient method until the forces acting on each atom were less than 0.01 eV/Å. The convergence criterion for self-consistent field calculations in energy was chosen of $10^{-4}$ eV. A $1 \times 8 \times 1$ k-point mesh was used to sample the Brillouin zone.

**Experimental results**

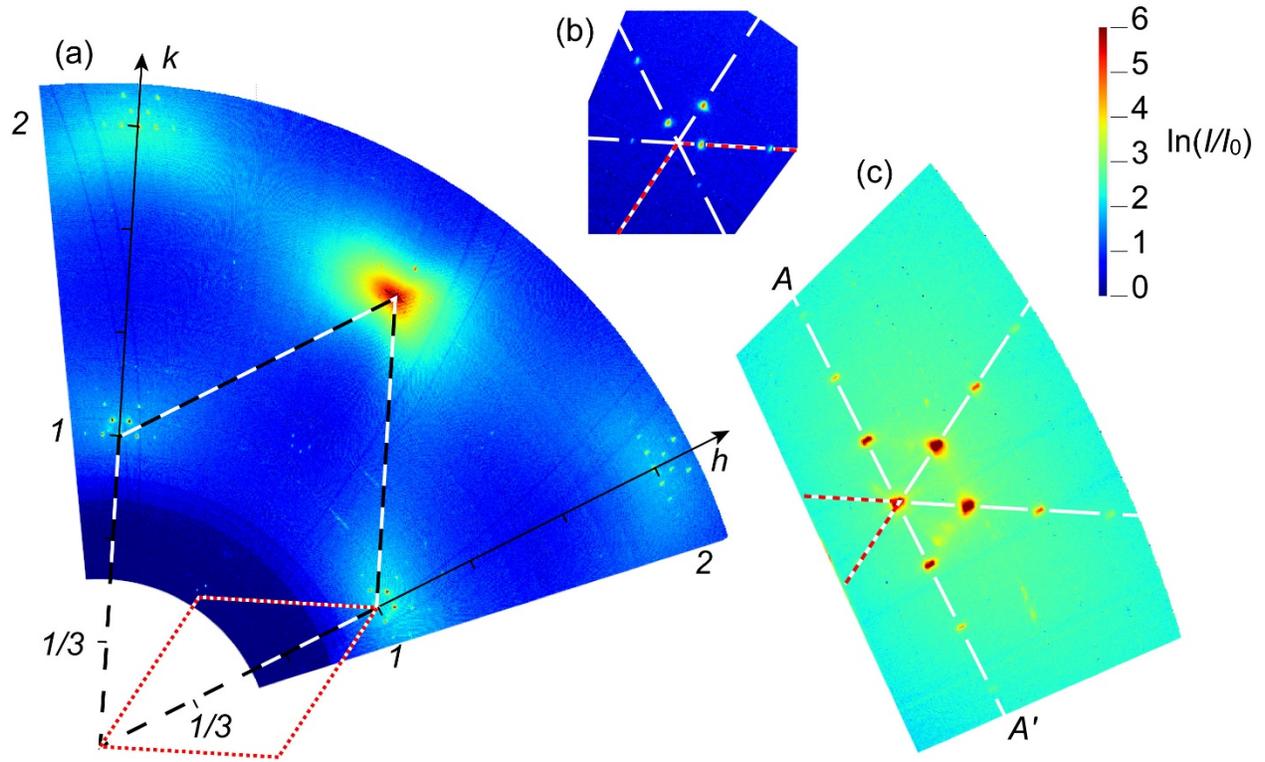

Fig. 1. Diffracted intensity for in-plane conditions ($l = 0.12$) after evaporation of $\theta_{\text{Ge}} \approx 1/3$ ML. (a) large view of the reciprocal space. The black dashed parallelogram corresponds to the Ag(111) surface unit cell. The red dotted parallelogram corresponds to a $(\sqrt{3} \times \sqrt{3})R30°$ supercell.

(b) detailed view around $(h = \frac{1}{3}, k = \frac{1}{3})$ condition. (c) detailed view around $(h = 1, k = 0)$ condition. The white dotted lines correspond to the directions of spot alignments.

Figure 1a shows the diffracted intensity for in-plane conditions ($l = 0.12$) after evaporation of $\theta_{Ge} \approx 1/3$ ML at T = 420 K. Satellite spots appear around diffraction spots corresponding to the crystal truncation rods (CTR) of the substrate (integer values of $h$ and $k$ indices). Other sets of spots appear around fractional values of ($h, k$), i.e. diffraction conditions corresponding to a ($\sqrt{3} \times \sqrt{3}$) reconstruction (for example, $h = \frac{1}{3}, k = \frac{1}{3}$). A detailed view of the diffracted intensity around ($\frac{1}{3}, \frac{1}{3}, 0.12$) and (1, 0, 0.12) is shown in Fig. 1b and 1c.

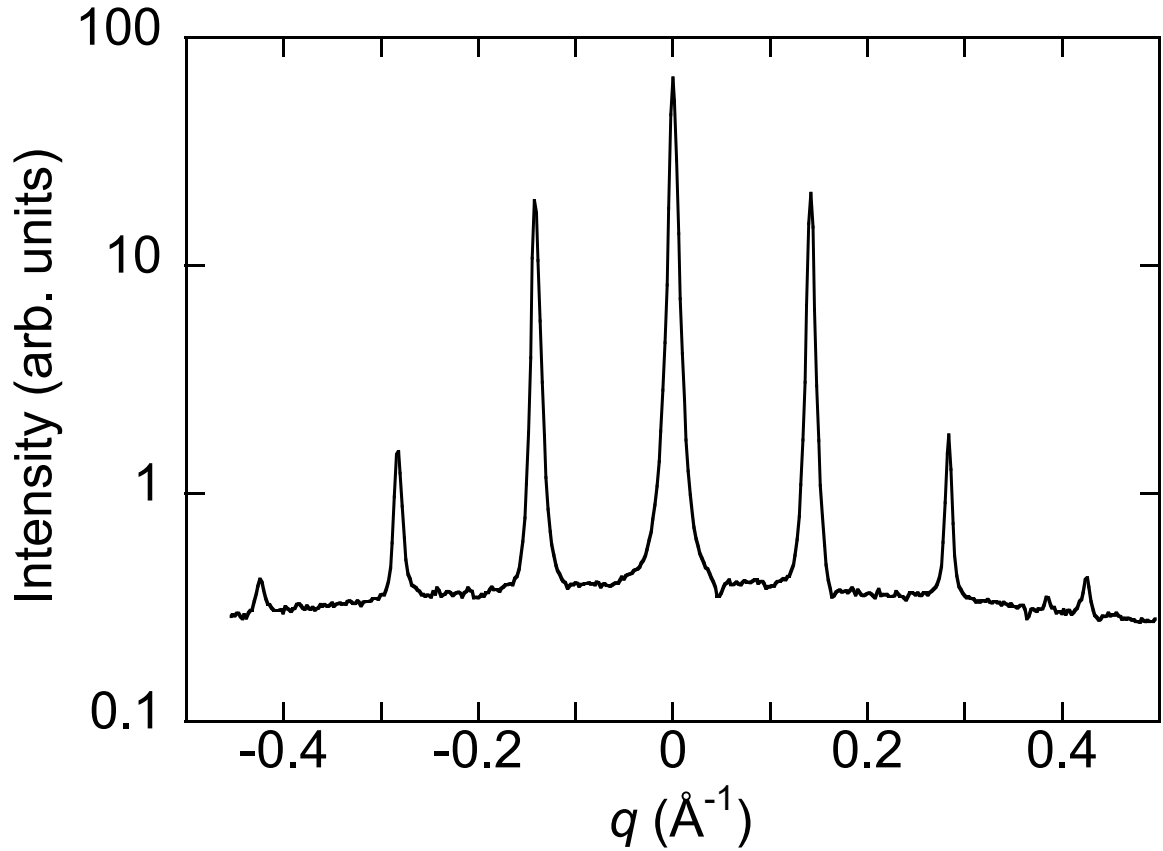

Fig. 2. Variation of the diffracted intensity for in-plane conditions around the ($h = 1, k = 0, l = 0.12$) position, corresponding to a scan along the A-A' line drawn on Fig. 1c.

These spots are aligned along three directions, indicated by white dotted lines in Fig. 1b and 1c. These <110> directions are equivalent due to the p3m1 symmetry of the substrate surface. Thus, the reconstruction should correspond to the three possible orientations of a similar reconstruction. We can exclude the fact that all spots belong to a single domain of trigonal or hexagonal symmetry since it would imply a large number of extinctions which are not observed in the present experiments.

A profile along the A-A' line shown in Fig. 1c is drawn in Fig. 2. The central spot corresponds to the CTR and 3 orders of diffraction from each side are visible for the satellite spots. The spacing between the spots is $\Delta q = 0.1412 \pm 0.0002$ Å$^{-1}$, indicating a periodicity of 44.5 Å, i.e, 15.4 times the Ag-Ag interatomic distance $a_{Ag} = 2.889$ Å. Thus, as expected from previous works [8], the measured phase appears as a periodic modulation of a ($\sqrt{3} \times \sqrt{3}$) reconstruction, along the <110> directions. A precise analysis of the diffraction diagram show that the unit cell is a rectangular centered unit cell. As 15.4 is close to 31/2, it can be described as a rectangular c(31 × $\sqrt{3}$) reconstruction. The value of $15.4 a_{Ag}$ is in good agreement with the periodicity of the striped pattern previously observed in STM images [10]. Thus, the diffraction pattern measured by SXRD corresponds to the striped pattern previously observed on STM images. On the contrary, we can exclude that this structure corresponds to a $(19\sqrt{3} \times 19\sqrt{3})R30°$ reconstruction as proposed in ref. [11].

The intensity and the shape of the spots in Fig. 1b and 1c are modulated by the elongated X-ray beam footprint on the sample, in the real space, and by the intrinsic width of the diffraction pattern, in the reciprocal space. For spots around CTRs, the intrinsic width is mainly given by the finite size of the striped domains and by the fluctuation of the striped phase periodicity. The first contribution is the same for all the spots whereas the second one increases linearly with the distance of the satellite spot to the substrate spot. The width of the satellite spots is measured to $\Delta q = 1.2\,10^{-3} + 6\,10^{-4} n$ Å$^{-1}$ where $n$ is the diffraction order. This shows that the periodicity is very well defined, and the lateral size of the domains is of the order of 0.5 µm.

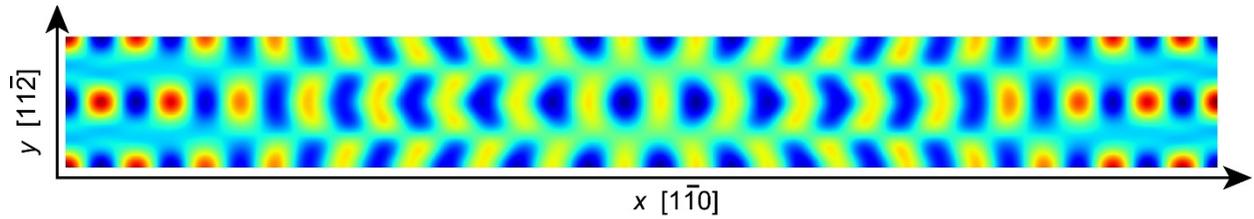

Fig. 3. Experimental Patterson map of the $c(31 \times \sqrt{3})$ structure. As it is centered, only half of the unit cell has been drawn along $x$. (size of the Patterson map is $44.78 \times 5.00$ Å$^2$).

From the measured in-plane structure factors, we have computed the 2D Patterson map, shown in Figure 3, which is an approximation of the electron density autocorrelation function within the surface unit cell [21]: $P(x,y) = 2\sum_{HK}|F(H,K)|^2 \cos(2\pi(Hx + Ky))$, where the $(H, K, L)$ indices refer to the $c(31 \times \sqrt{3})$ basis and $F(H,K)$ are the structure factors specific of this reconstruction, and measured for $L = 0.12$. Close to the origin ($x \approx 0$) and to the center of the unit cell ($x \approx 0.5$), a nearly perfect hexagonal lattice is visible, corresponding to a slightly contracted $(1 \times 1)$ unit cell with respect to the Ag(111) unit cell. It is worthwhile to notice that the modulation of the electron density autocorrelation associated with the $(\sqrt{3} \times \sqrt{3})$ local ordering is hardly visible in the Patterson map. This is explained by the intensity of the satellite spots near fractional values of $(h, k)$, much weaker than the intensity of the satellite spots near integer values of $(h, k)$ (see Fig. 1). This observation is in good agreement with a chemical ordering between Ag and Ge atoms. Indeed, if one third of the surface atoms are Ge atoms, the intensity associated with $(\sqrt{3} \times \sqrt{3})$ satellites should roughly scale as $\left(Z_{Ge} + 2Z_{Ag}\cos\left(\frac{2\pi}{3}\right)\right)^2 \approx 225$ whereas intensity associated with $(1 \times 1)$ satellites should roughly scale as $\left(Z_{Ge} + 2Z_{Ag}\right)^2 \approx 15900$. The contribution of the chemical ordering to the autocorrelation function is thus only of the order of 1%. Note that in a model of germanene, with only Ge atoms in a honeycomb lattice, this intensity ratio would be much higher, namely $\left(\cos\left(\frac{2\pi}{3}\right)\right)^2 = \frac{1}{4}$, which is not in agreement with the experimental observations of Fig. 1.

In the experimental Patterson map, along the long side of the $c(31 \times \sqrt{3})$ unit cell, corresponding to the $[1\bar{1}0]$ direction, a series of 33 maxima of intensity is visible. In between the node and the center of the unit cell ($x \approx 0.25$ or $0.75$), the maxima of correlations appear as elongated along $y$ ($[11\bar{2}]$ direction). This indicates that atoms are regularly spaced along the $[1\bar{1}0]$ direction, but at slightly different positions along the $[11\bar{2}]$ direction.

From these observations, we propose that the striped phase has a density 33/31 times higher than the Ag(111) atomic density, and that the atomic positions undulate between the fcc sites and the hcp sites. Such structure would thus display a great analogy with the $(22 \times \sqrt{3})$ Au(111) reconstruction for which the surface atomic density is 23/22 times higher than the one of a Au(111) bulk plane [22]. Consequently, starting from this Ag$_2$Ge alloy surface model, we have relaxed their atomic positions by DFT, and we have checked the validity of the relaxed model by comparing the theoretical structure factors corresponding to the model, to the experimental ones.

**DFT computations**

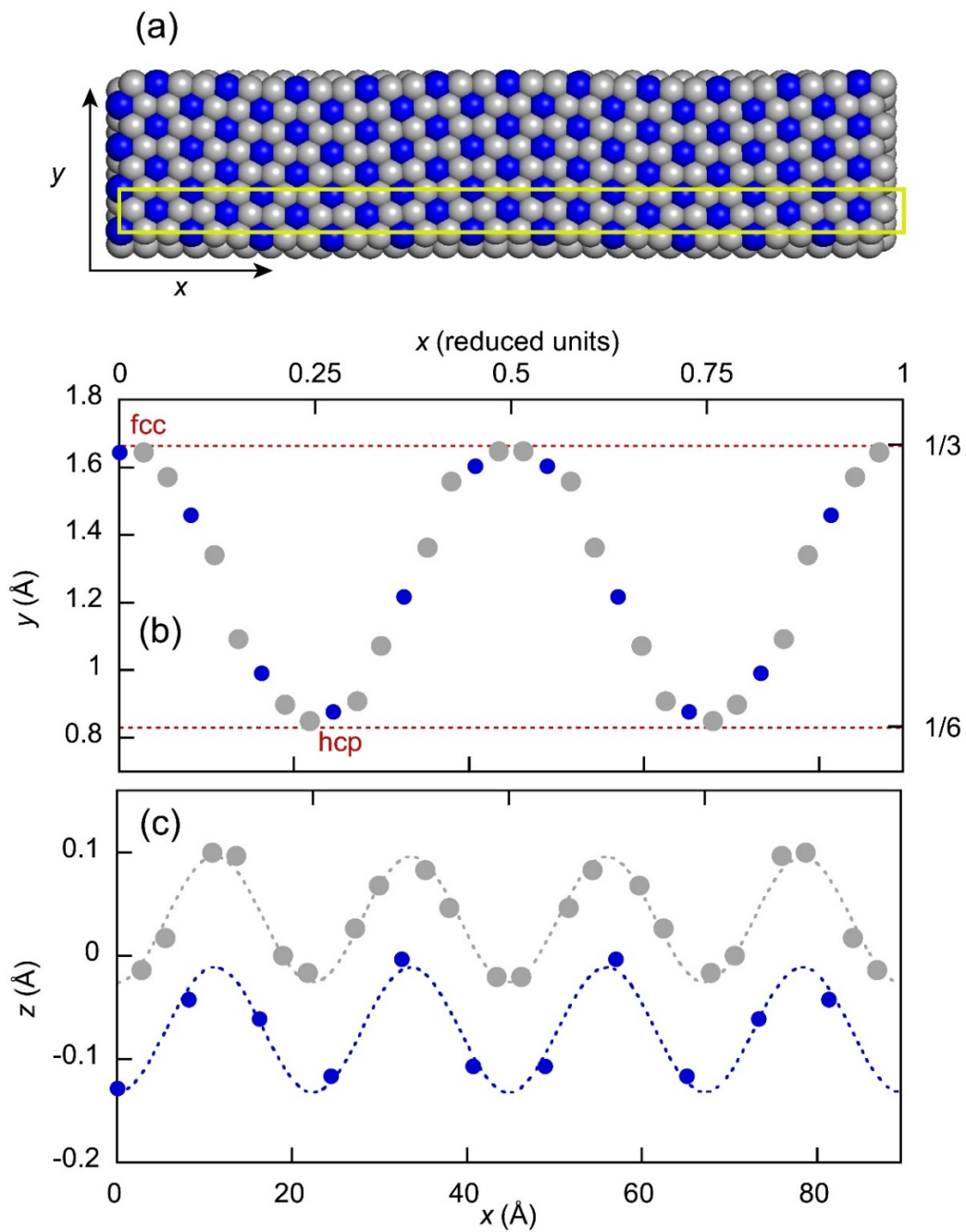

Fig. 4. Top view (a) and profiles (b, c) along $x$ for surface atoms (blue: Ge, grey: Ag). The scales along $x$ and $y$ directions are given both in reduced units and in Å. The $c(31 \times \sqrt{3})$ unit cell is drawn in yellow.

Figure 4 shows the configuration for the alloyed structure obtained after relaxation. The $y(x)$ profile indicates that the atomic positions of surface atoms oscillate between fcc (at $y = 1/3$) and hcp sites (at $y = 1/6$). The $z(x)$ profiles in Fig. 4c shows that on average, Ge atoms are located around 0.1Å below the Ag atoms. The $z(x)$ profiles for Ge and Ag atoms display a periodic oscillation with a double frequency as compared with $y(x)$. Starting from 0, the odd and even minima for $z$ correspond to atoms in fcc and hcp positions respectively, whereas maxima correspond to atoms in bridge position. This gives rise to the periodic striped pattern observed in STM images [8]. The calculations indeed indicate an apparent periodicity of 22.4 Å with to 5.5 Ge atoms per stripe, in good agreement with the $6\sqrt{3}$ periodicity found by STM [8]. The value of the buckling is 0.12Å, whereas the buckling measured by STM is approximately 0.2 Å [8]. The Ge-Ag interatomic distances are slightly smaller for Ge atoms in bridge position (2.662 Å) than for Ge atoms in fcc position (2.722 Å). The value of the relaxed atomic positions is given in the Supplemental Material [23]. This undulation of the atomic positions between the fcc sites and the hcp sites is strongly analogue to the $(22 \times \sqrt{3})$ Au(111) reconstruction, for which the surface atomic density is 23/22 times higher than the one of a Au(111) bulk plane, and which displays also a similar striped structure [22].

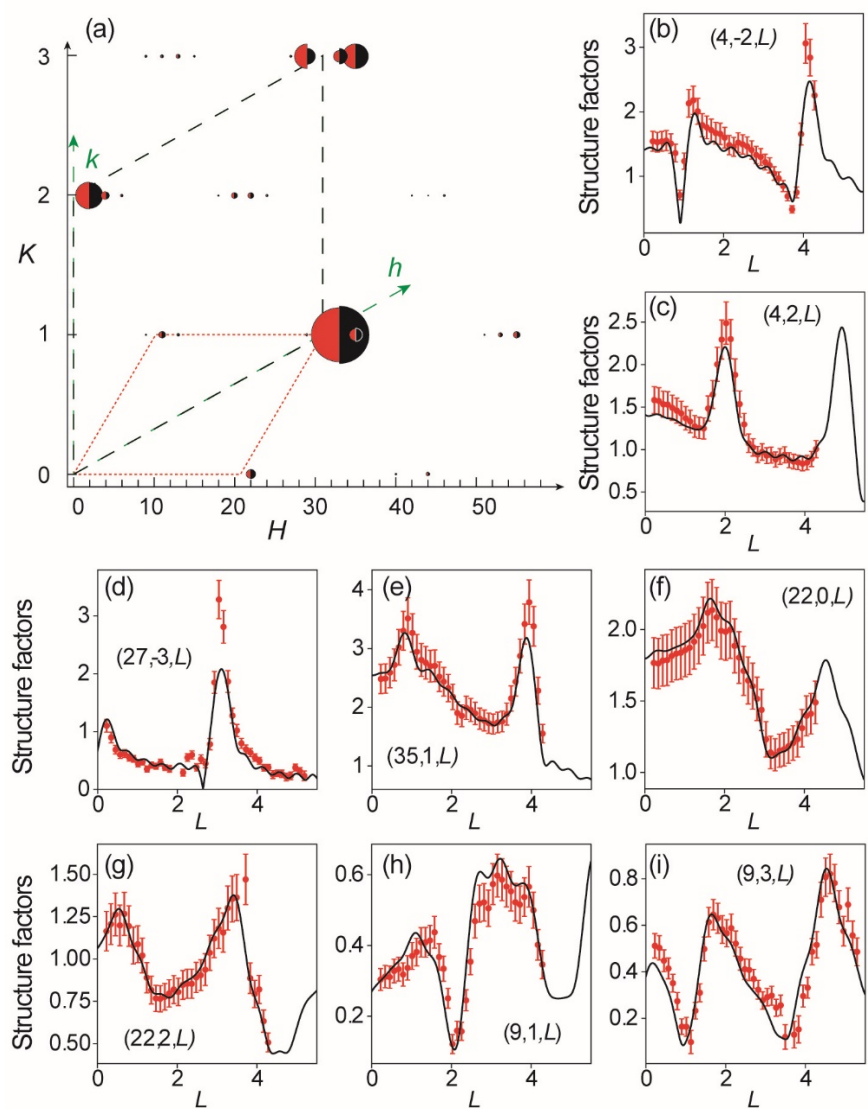

Fig. 5. (a) Comparison between experimental (red half-disks) and simulated (black half-disks) in-plane structure factors. The black dashed parallelogram corresponds to the Ag(111) surface unit cell. The substrate structure factors for integer values of $h$ and $k$ have not been drawn for clarity. The red dotted parallelogram corresponds to a $(\sqrt{3} \times \sqrt{3})R30°$ supercell. (b-i) Comparison between experimental (red dots) and simulated (black line) structure factors along $(H, K, L)$ satellite rods, near integer values of $(h, k)$ (b-e) or near fractional values of $(h, k)$ (f-i).

**Comparison with SXRD and discussion**

From this atomic configuration relaxed by DFT, we have computed the simulated structure factors and compared them to the experimental ones. Only two scale factors (one for the in-plane set and one for the rods) and Debye-Waller (DW) factors along the three directions were used as free parameters. For simplicity, we have set identical DW factors for all Ge atoms, for all Ag surface atoms, for all Ag atoms of the second plane and for all other Ag atoms. Thus only 14 free parameters were used to fit the data. The agreement between experimental and simulated ($F_{th}$) structure factors is estimated by the value of $\chi^2 = \frac{1}{N_{pts}-N_{par}}\Sigma_{N_{Pts}}\left(\frac{F_{th}-F_{exp}}{\sigma_{exp}}\right)^2$ where $N_{pts} = 2493$ is the number of experimental structure factors, $N_{par}$ is the number of free parameters and $\sigma_{exp}$ is the experimental uncertainty, which takes into account the statistical uncertainty given by the number of counted photons and an overall 10% uncertainty estimated from the comparison of symmetry-related structure factors.

Fig. 5 shows the comparison between the experimental and simulated structure factors $F$ for in-plane measurements and along selected rods. Since $F(H, K, 0) = F(-H, -K, 0)$ and $F(H, K, 0) = F(-H, K, 0)$ due to the mirror symmetry of the unit cell, only the quadrant corresponding to $H > 0, K > 0$ has been drawn. The complete comparison of 62 non-equivalent reconstruction rods is given in Fig. S1 and the Debye Waller parameters used for the fit are given in Table S1 [23]. There is an outstanding agreement between experiments and simulations, corresponding to a small value $\chi^2 = 1.90$. As can be seen, the simulation reproduces very well both satellite rods near integer values of $(h, k)$ (Fig. 5b-e) and satellite rods near fractional values of $(h, k)$ (Fig. 5f-i). The first ones display intense variations at specific integer values of $L$, for example, near $L = 1$ and near $L = 4$ in Fig. 5b and 5e, near $L = 2$ in Fig. 5c, or near $L = 0$ and near $L = 3$ in Fig. 5d. These positions are close to the Bragg diffraction conditions of the Ag(111) crystal. These intense variations can be attributed to the periodic elastic relaxations that penetrate in the bulk [24,25]. Such relaxations are induced by the 6.45% misfit between the outermost surface layer and the substrate. Thus, the fact that they are nicely reproduced indicates that DFT simulations give a very accurate description of the interaction between the surface layer and the substrate, even if the number of layers free to relax has been limited to 7 in the DFT calculations.

In order to estimate the uncertainties related to the atomic positions, we have let all atomic positions free to move in the first two planes, starting from the values obtained by DFT. This leads to a huge number of free parameters ($N_{par}$=111) and to a reduction of $\chi^2$ down to 0.9. The atomic positions are not strongly modified. Displacements from the initial positions are less than 0.2 Å for Ag and Ge atoms in the first layer and less than 0.1 Å for atoms in the second layer, with rms displacements of 0.08 Å, 0.08 Å and 0.05 Å in the first layer along the $x$, $y$, and $z$ directions respectively. Similar values are 0.03 Å, 0.03 Å, 0.06 Å for atoms in the second plane.

For the sake of completeness, we have also compared the experimental structure factors to the simulated structure factors obtained within a model of a germanene epitaxial layer. For that purpose, a model with 22 Ge hexagons along the long side of the mesh has been used, so that the Ge atomic density is twice the one of the alloyed model. The atomic positions obtained after relaxation by DFT are drawn in Fig. S2. As previously, we have set identical DW factors for all Ge atoms, for all Ag atoms of the second plane and for all other Ag atoms. We find a strong deviation between experiments and simulations as $\chi^2 = 26.5$. Therefore, the germanene layer cannot account the experimental observation and this model must clearly be rejected.

We have also computed the adsorption energy for Ge atoms in the two models:

$$E_{ad} = (E_{Ge-Ag} - E_{Ag} - \Delta N_{Ag} E_{Ag\,bulk})/N_{Ge}$$

Where $E_{Ge-Ag}$ is the total energy of the considered model, $E_{Ag}$ is the energy of a slab of same lateral size without Ge, $E_{Ag\,bulk} = -2.72$ eV is the bulk cohesive energy of Ag atoms, $N_{Ge}$ the number of Ge atoms in the slab, and $\Delta N_{Ag}$ is the difference of number of Ag atoms between the considered Ge/Ag model and the bare Ag slab.

$E_{ad}$ is lower for the Ag$_2$Ge surface alloy (-4.56 eV) than for the germanene layer (-4.41 eV). Taking as a reference the Ge bulk cohesive energy (-4.45 eV), the germanene layer is less stable than the Ge bulk whereas the Ag$_2$Ge surface alloy is more stable. This indicates that, from a thermodynamic point of view, wetting of the Ag surface with a Ag$_2$Ge alloy is favored, whereas whereas a germanene layer would be unstable and should dewet to form 3D Ge crystallistes. Thus, there is a better thermodynamic stability for the Ag$_2$Ge surface alloy than for the germanene layer.

The atomic structure of the striped phase presents a great analogy with the $(22 \times \sqrt{3})$ Au(111) reconstruction. This reconstruction corresponds to a surface atomic density 23/22 times higher than the one of a Au(111) bulk plane [22]. For the Au surface atoms which are less coordinated than the bulk ones, the strength of the bonds is higher than in the bulk. Their equilibrium interatomic distance would thus be lower than in the bulk. This results in a large tensile stress for surface atoms. For Au(111), the system relaxes in order to reduce the interatomic distances for surface atoms, i.e., by increasing the surface atomic density. The energy gain due to the decay of the interatomic distances at the surface exceeds the energy cost of the regions where atoms occupy bridge positions instead of three-fold coordinated hcp or fcc positions. Unlike Au(111), Ag(111) does not spontaneously reconstruct. However, when a Ge atom replaces an Ag atom of the surface, this increases the absolute value of the tensile surface stress since the atomic radius of Ge (1.25 Å) is less than the one of Ag (1.60 Å). Above a critical coverage, the system relaxes in a configuration where the interatomic distances are decreased, at the cost of the creation of Shockley partial dislocations, i.e., discommensuration lines separating regions where atoms occupy fcc positions, from regions where atoms occupy hcp positions. Other systems have been reported to show a similar behavior such as Cu, Ag, Au films on Ru(0001) [26,27].

**Conclusion**

Using GIXD, we have unambiguously determined the structure of the striped pattern observed upon Ge deposition on Ag(111) at 420 K, which was controversially attributed to a pure germanene layer or to a Ge-Ag alloy. From a careful analysis of the Patterson map, leading to an initial structural model of the system, we obtained the model of the $Ag_2Ge$ surface alloy with an atomic density 6.45% higher than the one of Ag(111). The computed theoretical structure factors obtained from this model, relaxed by DFT, show an outstanding agreement with the experimental ones, demonstrating that this striped phase is actually a $Ag_2Ge$ surface alloy. The observed stripes correspond to Shockley partial dislocations that separate alternating fcc and hcp domains, presenting a strong analogy with the Au $(22 \times \sqrt{3})$ reconstruction. A model of germanene can be ruled out since it does not fit the experiments and since it is thermodynamically less stable.

It has to be noted that such a determination was possible thanks to the very large set of acquired structure factors, which can now easily be done within a relatively short time thanks to 2D detectors. This shows how powerful DFT associated with GIXD can be for the analysis of complex surface structures. In particular, since DFT is shown here to give a very accurate description of the atomic relaxations, the comparison between GIXD and DFT is a criterium of choice for determining the validity of an atomic model of a surface structure.

**Acknowledgments**

This study is financially supported the French National Research Agency (Germanene project ANR-17-CE09-0021-03). K.Z. is supported by the Chinese Scholarship Council (CSC contract 201808070070).

# Resolving the structure of the striped Ge layer on Ag(111): Ag$_2$Ge surface alloy with alternate fcc and hcp domains

K. Zhang,[1] D. Sciacca,[2] R. Bernard,[1] Y. Borensztein,[1] A. Coati,[3] P. Diener,[2] B. Grandidier,[2] I. Lefebvre,[2] M. Derivaz,[4] C. Pirri,[4] G. Prévot[1]

[1] Sorbonne Université, Centre National de la Recherche Scientifique, Institut des NanoSciences de Paris, INSP, F-75005 Paris, France

[2] Univ. Lille, CNRS, Centrale Lille, Univ. Polytechnique Hauts-de-France, Junia-ISEN, UMR 8520 - IEMN, F-59000 Lille, France

[3] Synchrotron SOLEIL, L'Orme des Merisiers Saint-Aubin, BP 48 91192 Gif-sur-Yvette Cedex, France

[4] Institut de Science des Matériaux de Mulhouse IS2M UMR 7361 CNRS-Université de Haute Alsace, 3 bis rue Alfred Werner, 68057, Mulhouse, France

**Supplemental Material**

|  | along *x* | along *y* | along *z* |
|---|---|---|---|
| Ge | 0.46 | 1.19 | 3.16 |
| Ag surface | 1.46 | 1.73 | 3.75 |
| Ag second plane | 0.52 | 1.90 | 4.85 |
| Other Ag | 0 | 2.00 | 3.60 |

Table S1. Debye-Waller parameters ($\text{Å}^2$) used for fitting the experimental structure factors.

| atom | x | y | z |
|---|---|---|---|
| Ag | 0.0000 | 0.0036490 | 1.6505 |
| Ag | 0.032279 | 0.0033225 | 1.6510 |
| Ag | 0.064590 | 0.0023605 | 1.6532 |
| Ag | 0.096809 | 0.00059250 | 1.6550 |
| Ag | 0.12902 | 0.99914 | 1.6562 |
| Ag | 0.16126 | 0.99935 | 1.6557 |
| Ag | 0.19347 | 0.99996 | 1.6533 |
| Ag | 0.22577 | 0.99989 | 1.6522 |
| Ag | 0.016169 | 0.50367 | 1.6506 |
| Ag | 0.048434 | 0.50261 | 1.6519 |
| Ag | 0.080681 | 0.50132 | 1.6540 |
| Ag | 0.11294 | 0.50007 | 1.6562 |
| Ag | 0.14512 | 0.49923 | 1.6560 |
| Ag | 0.17735 | 0.49929 | 1.6546 |
| Ag | 0.20964 | 0.50030 | 1.6528 |
| Ag | 0.24191 | 0.50058 | 1.6516 |
| Ag | 0.016129 | 0.17149 | 1.9831 |
| Ag | 0.048401 | 0.17029 | 1.9843 |
| Ag | 0.080699 | 0.16877 | 1.9884 |
| Ag | 0.11289 | 0.16671 | 1.9899 |
| Ag | 0.14508 | 0.16577 | 1.9899 |
| Ag | 0.17735 | 0.16586 | 1.9895 |
| Ag | 0.20960 | 0.16622 | 1.9847 |
| Ag | 0.24192 | 0.16649 | 1.9833 |
| Ag | 0.0000 | 0.67131 | 1.9824 |
| Ag | 0.032300 | 0.67119 | 1.9839 |
| Ag | 0.064540 | 0.66960 | 1.9860 |
| Ag | 0.096760 | 0.66771 | 1.9886 |
| Ag | 0.12902 | 0.66629 | 1.9919 |
| Ag | 0.16120 | 0.66561 | 1.9890 |
| Ag | 0.19346 | 0.66614 | 1.9859 |
| Ag | 0.22577 | 0.66636 | 1.9860 |
| Ag | 0.0000 | 0.33918 | 2.3146 |
| Ag | 0.032260 | 0.33865 | 2.3165 |
| Ag | 0.064449 | 0.33817 | 2.3195 |
| Ag | 0.096719 | 0.33536 | 2.3224 |
| Ag | 0.12904 | 0.33129 | 2.3256 |
| Ag | 0.16115 | 0.33342 | 2.3237 |
| Ag | 0.19354 | 0.33187 | 2.3200 |
| Ag | 0.22579 | 0.32893 | 2.3168 |
| Ag | 0.016090 | 0.83943 | 2.3157 |
| Ag | 0.048350 | 0.83828 | 2.3168 |
| Ag | 0.080641 | 0.83536 | 2.3219 |
| Ag | 0.11280 | 0.83568 | 2.3245 |
| Ag | 0.14511 | 0.83254 | 2.3247 |
| Ag | 0.17740 | 0.82934 | 2.3218 |
| Ag | 0.20956 | 0.83249 | 2.3185 |
| Ag | 0.24199 | 0.83200 | 2.3167 |
| Ag | 0.0000 | 0.011352 | 2.6495 |
| Ag | 0.032240 | 0.0065685 | 2.6492 |
| Ag | 0.064100 | 0.0046095 | 2.6484 |
| Ag | 0.096430 | 0.0051580 | 2.6595 |
| Ag | 0.12902 | 0.99613 | 2.6626 |
| Ag | 0.16136 | 0.99641 | 2.6509 |
| Ag | 0.19351 | 0.99518 | 2.6544 |
| Ag | 0.22590 | 0.99051 | 2.6511 |
| Ag | 0.015879 | 0.50728 | 2.6466 |
| Ag | 0.048099 | 0.50958 | 2.6520 |
| Ag | 0.080509 | 0.50247 | 2.6570 |
| Ag | 0.11263 | 0.49971 | 2.6515 |
| Ag | 0.14500 | 0.49946 | 2.6623 |
| Ag | 0.17758 | 0.49246 | 2.6577 |
| Ag | 0.20988 | 0.49399 | 2.6477 |
| Ag | 0.24183 | 0.49336 | 2.6482 |
| Ge | 0.0000 | 0.32894 | 2.9778 |
| Ge | 0.18167 | 0.19843 | 2.9873 |
| Ge | 0.091420 | 0.29182 | 2.9900 |
| Ge | 0.045889 | 0.82069 | 2.9808 |
| Ge | 0.13655 | 0.74365 | 2.9955 |
| Ge | 0.22715 | 0.67551 | 2.9795 |
| Ag | 0.030400 | 0.32873 | 2.9940 |
| Ag | 0.061410 | 0.31421 | 2.9984 |
| Ag | 0.12124 | 0.26822 | 3.0101 |
| Ag | 0.15187 | 0.21862 | 3.0096 |
| Ag | 0.21168 | 0.17960 | 2.9960 |
| Ag | 0.24261 | 0.17013 | 2.9936 |
| Ag | 0.015550 | 0.82946 | 2.9931 |
| Ag | 0.076069 | 0.81177 | 3.0025 |
| Ag | 0.10683 | 0.77247 | 3.0077 |
| Ag | 0.16629 | 0.71457 | 3.0056 |
| Ag | 0.19702 | 0.68168 | 2.9997 |

Table S2. Atomic positions near the surface given in reduced units (unit cell dimensions: $a = 89.545$ Å, $b = 5.003$ Å, $c = 7.075$ Å, $\alpha = \beta = \gamma = 90°$) for the Ag$_2$Ge surface alloy. Only 1/4$^{th}$ of the unit cell is given, the other positions are obtained with the transformations $x' = -x$ and $(x', y') = (x + 1/2, y + 1/2)$.

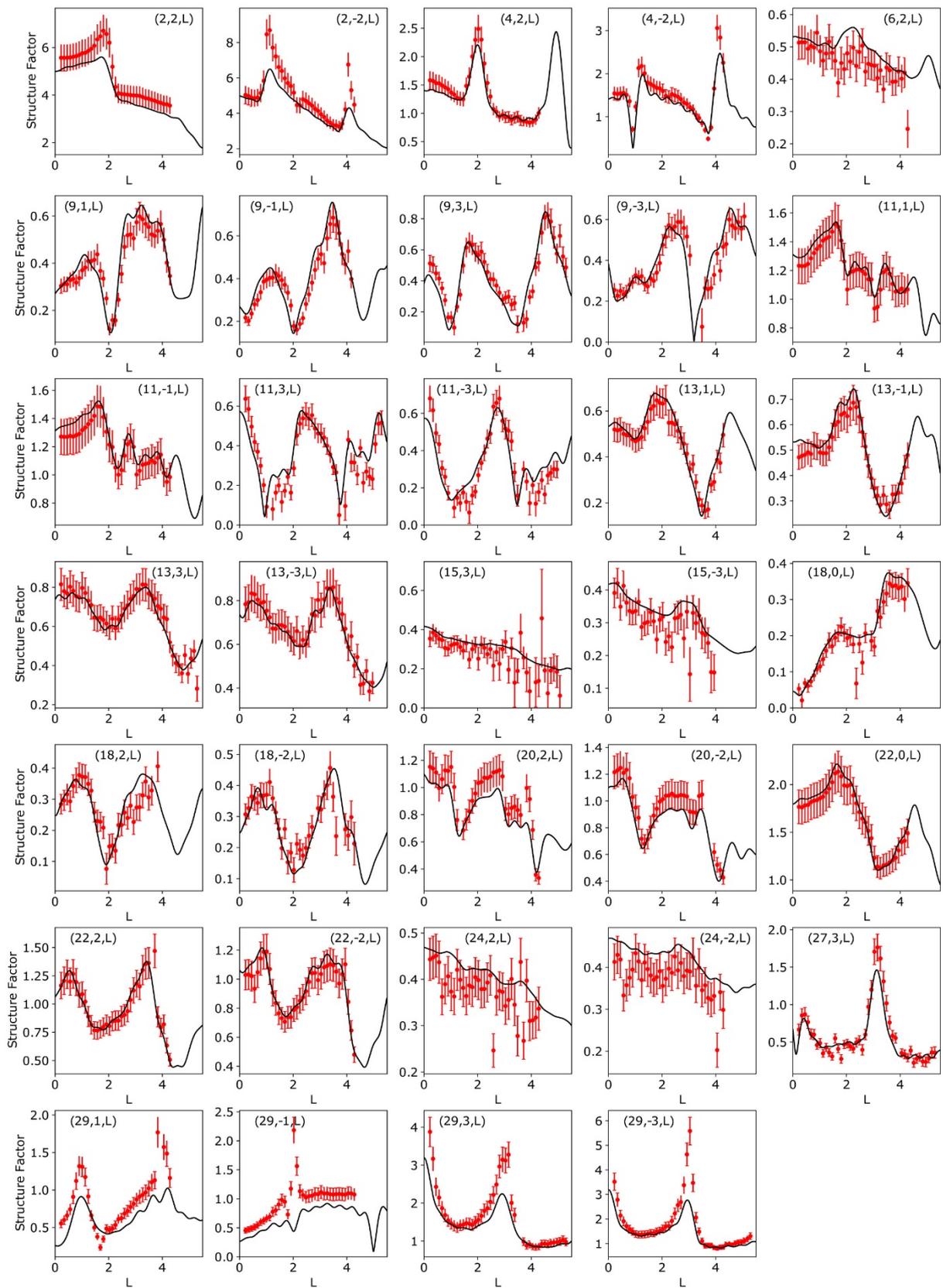

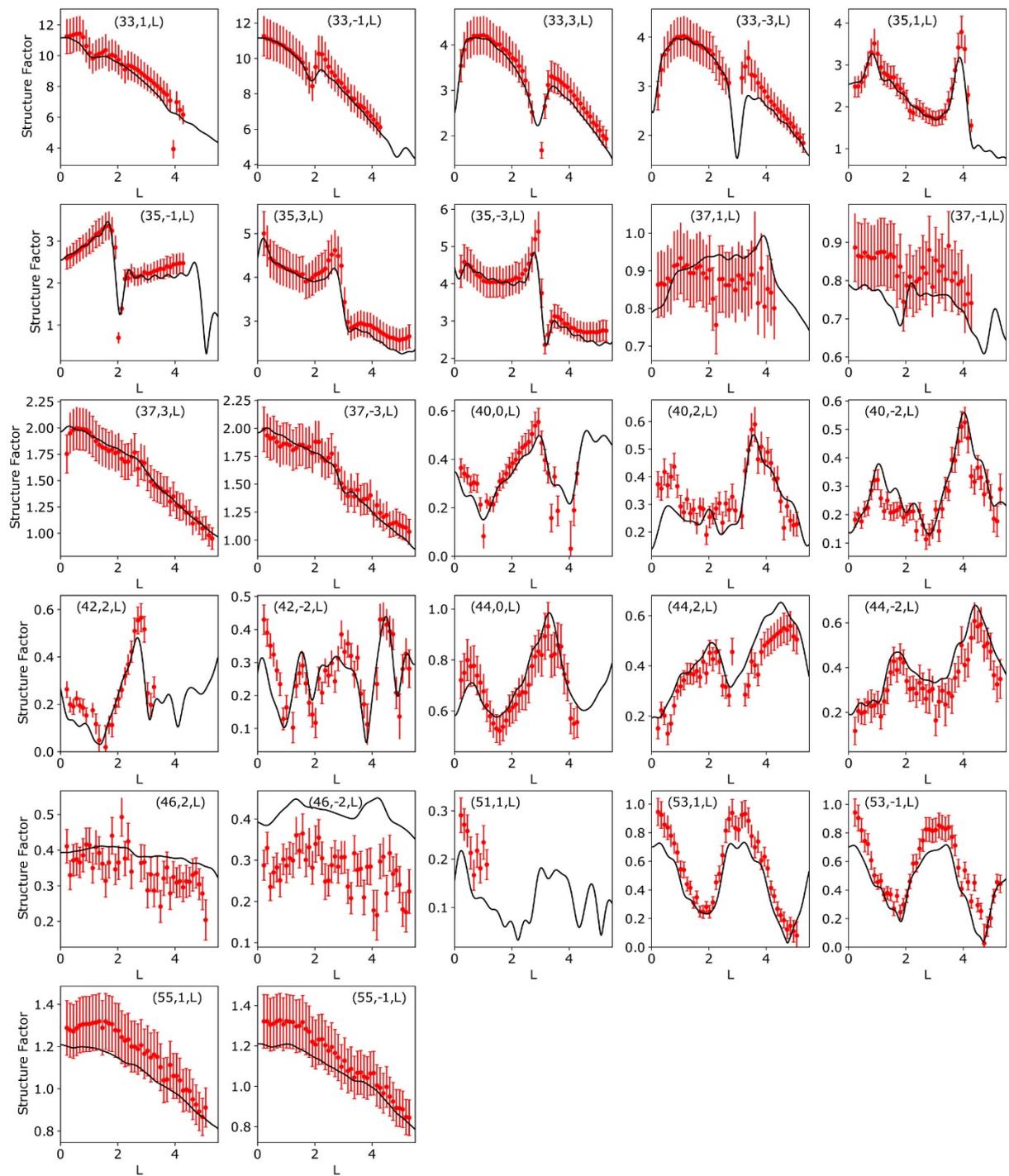

Fig. S1. Comparison between experimental (red dots) and simulated (black line) structure factors along all measured rods.

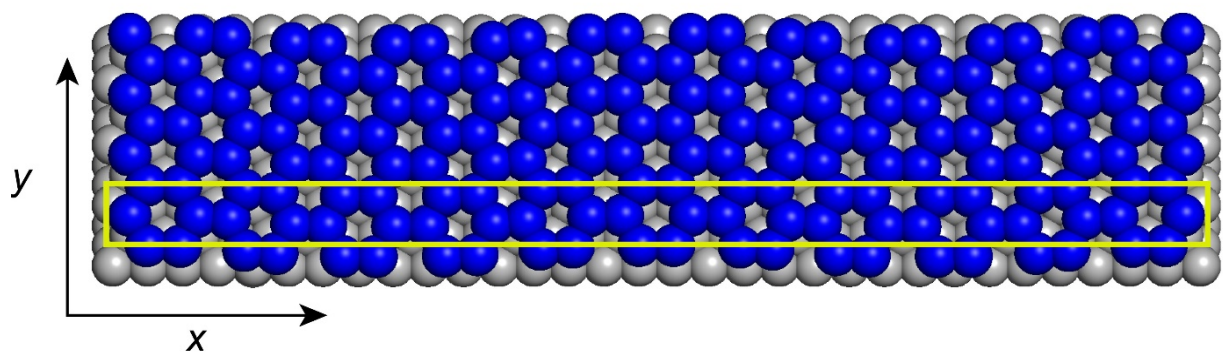

Fig. S2. Top view of a germanene layer on Ag(111). The c(31 × √3̄) unit cell is drawn in yellow.